\documentclass[prl,twocolumn,floatfix,showpacs]{revtex4}
\usepackage{amssymb}
\usepackage{amsmath}
\usepackage{epsfig}
\usepackage{graphicx}

\begin{document}

\title { Coincidence of the oscillations in the dipole transition and
 in the persistent current of narrow quantum rings with two electrons }

\author {Y. Z. He}
\author {C. G. Bao$^{\ast }$ }

\affiliation{State Key Laboratory of Optoelectronic Materials and
Technologies, and Department of Physics, Sun Yat-Sen University, \
Guangzhou, 510275, P.R. China}

\begin{abstract}
The fractional Aharonov-Bohm oscillation (FABO) of narrow quantum
rings with two electrons has been studied and has been explained
in an analytical way, the evolution of the period and amplitudes
against the magnetic field  can be exactly described. Furthermore,
the dipole transition of the ground state was found to have
essentially two frequencies, their difference appears as an
oscillation matching the oscillation of the persistent current
exactly.  A number of equalities relating the observables and
dynamical parameters have been found.
\end{abstract}

 \maketitle

* The corresponding author

\vspace{1pt}

Quantum rings containing only a few electrons can be now
fabricated in laboratories$^{1,2}.$  When a magnetic field $B$\ is
applied, interesting physical phenomena, e.g., Aharonov-Bohm
oscillation (ABO) and fractional ABO (FABO)of
the ground state (GS) energy $E_o$ and persistent current $J_o$, have been observed $%
^{2-4,13}$. In the theoretical aspect, a number of calculations
based on exact diagonalization$^{5-8}$, local-spin-density
approximation$^{9,10}$, and the diffusion Monte Carlo
method$^{11}$ have been performed.  These calculations can in
general reproduce the experimental data. \ For examples, in the calculation of 4-electron ring%
$^{6,11}$, the period of oscillation \ $\Phi _{0}/4$ found in
experiments was recovered  $(\Phi _{0}=hc/e is the flux quantum)$.

In addition to the oscillations in $E_o$ and $J_o$, the
oscillation in the optical properties is noticeable.$^{16,17}$. In
this paper a new kind of oscillation found in the dipole
transition of two-electron (2-e) narrow rings is reported. The
emitted (absorbed) photon of the dipole transition of the GS was
found to have essentially two energies, their difference is
exactly equal to $hJ_o$, where $h$ is the Planck's constant. In
other words the difference of the two photon energies appears as
an oscillation which matches exactly the oscillation of $J_o$.
This finding is approved by both numerical calculation and
analytical analysis as follows.

  The narrow 2-e ring is first considered as one-dimensional, then the effect of the width of the ring
is further evaluated afterward. The Hamiltonian reads

\ \ \

\qquad \qquad $H=T+V_{12}+H_{Zeeman}\qquad \qquad (1)$

\ \ \

\qquad \qquad $T=\sum\limits_{j=1}^{N}G(-i\frac{\partial }{
\partial \theta _{j}}+\Phi )^{2},\;\;\;G=\frac{\hbar ^{2}}{2m^{\ast }R^{2}}$

\ \ \

where $m^{\ast }$ the effective mass, $\theta _{j}$ the
azimuthal angle of the $j-th$ electron, \ $\Phi =\pi R^{2}B/\Phi _{0}$, where $B$ is
a magnetic field perpendicular to the plane of the ring,  $%
V_{12}$\ the e-e Coulomb interaction, \ $H_{Zeeman}=-S_{Z}\mu \Phi $ \ the
well known Zeeman energy where $S_{Z}$ is the Z-component of the total spin $%
S$, and $\mu =\frac{g^{\ast } \mu _{B}}{\pi R^2 \Phi _{0}}$, where
$g^{\ast }$ is the effective g-factor and $\mu _{B}$ is the Bohr
magneton. \ The interaction is adjusted as $^{7}$
$V_{12}=e^{2}/({2\varepsilon \sqrt{d^{2}+R^{2}\sin ^{2}((\theta
_{1}-\theta _{2})/2)}} )^{-1}$, where $\varepsilon $ is the
dielectric constant and the parameter $d$ is introduced to account
for the effect of finite thickness of the ring.

We first perform a numerical calculation so that all related
quantities can be evaluated
quantitatively. \ $m^{\ast }=0.063m_{e}$, $\varepsilon =12.4$ (for InGaAs), $%
d=0.05R$ , and the units $meV$, $nm$\ , $Tesla$ and $\Phi _{0}$ are used. \
Accordingly, $G=604.8/R^{2},$and$\;$\ $\mu =33.53/R^{2}$.\vspace{1pt}

 A set of basis functions $\phi _{k_{1}k_{2}}=e^{i(k_{1}\theta
_{1}+k_{2}\theta _{2})}/2\pi $\ is introduced to diagonalize the
Hamiltonian, where $k_{1}$ and $k_{2}$ must be integers to assure
the periodicity, the sum of $k_{1}$ and $k_{2}$ is just the total
orbital angular momentum $L$. \ $\phi _{k_{1}k_{2}}$ must be
further (anti-)symmetrized when $S=0(1)$. When about three
thousand basis functions are adopted, accurate solutions (at least
six effective digits) can be obtained.  The low-lying spectrum is
plotted in Fig.1, where the oscillation of the GS energy and the
transition of the GS angular momentum $L_o$ can be clearly seen.

 Let $\theta _{C}=(\theta _{2}+\theta _{1})/2$ , and $\varphi =\theta
_{2}-\theta _{1}$. \ Then

\ \ \ \

\vspace{1pt}\qquad $H=H_{coll}+H_{int}\qquad \qquad (2)$

\ \ \ \

where $H_{coll}=\frac{1}{2}G(-i\frac{\partial }{\partial \theta
_{C}}+2\Phi )^{2}+H_{Zeeman}$ and $H_{int}=2G(-i\frac{\partial
}{\partial \varphi })^{2}+V_{12}$, they are for the collective and
internal motions, respectively. \ Our numerical results lead to
the following points.

(i) \textit{Separability}: The separability of one-dimensional
ring is well known$^{5}$.  However, for the convenience of the
following description, it is briefly summarized as follows. Each
eigenenergy $E$\ can be exactly divided as a sum of three terms

\ \ \ \

 $E=\frac{1}{2}G(L+2\Phi )^{2}+E_{int}-S_{Z}\mu \Phi
\qquad \qquad (3)$

\ \ \ \

where the first term is the kinetic energy of collective motion, $E_{int}$
is the internal energy .

Since the basis functions can be rewritten as

\ \ \ \

  $\phi _{k_{1}k_{2}}=e^{iL\theta _{C}}e^{i\frac{1}{%
2}(k_{2}-k_{1})\varphi }/2\pi \qquad \qquad (4)$

\ \ \ \

 the spatial part of each eigenstate $\Psi $ is
strictly separable as  $\Psi =\frac{1}{\sqrt{2\pi }}e^{iL\;\theta
_{C}}\psi _{int}$  where the first part describes the collective
motion, \ while $\psi _{int}$ is a normalized internal state
depending only on $\varphi $.  In particular, both $E_{int}$ and
$\psi _{int}$
 do not depend on $B$ (or $\Phi $).

\begin{figure} [htbp]
\centering
\includegraphics[totalheight=2.2in,trim=30 40 5 10]{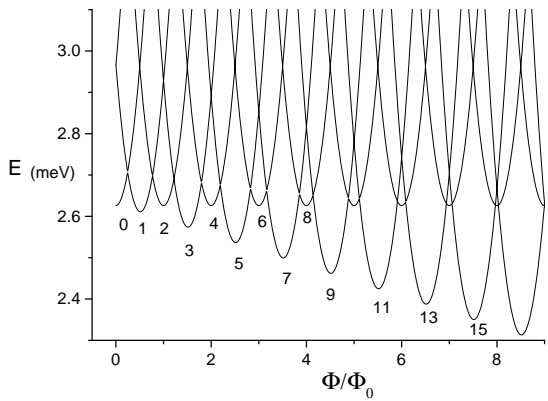}
\caption{ Low-lying levels of a 2-e ring against $\Phi /\Phi _{0}$
in the FABO\ region. \ When $\Phi /$ is positive, $L_{o}$ is
negative, the numbers by the curves are $-L_{o}$.} \label{Fig.1}
\end{figure}

(ii) \textit{Classification of $\psi _{int}$}:  When $L$ is even\
(odd), $(k_{2}-k_{1})/2$ is an integer (half-integer), thus the
period of $\varphi $ as shown in (4) is $2\pi \;(4\pi )$.
Therefore, the periodicity of the internal states have two
choices.  In fact, the difference in the periodicity is closely
related to the dependence of the domains of the new variables
$\theta _{C}$ and $\varphi$, this point has been discussed in
detail in ref.[14,15].  Let $Q=(-1)^{L}$, then the four cases
$(Q,S)=$ (1,0), (-1,0), (-1,1) and (1,1) are associated with four
types of states labeled by $a,\;b,\;c,$ and $d$ , respectively. \
\ The internal states of Type $a$ are denoted as $\psi _{a}$,
$\psi _{a^{\ast }}$,  $\cdot \cdot \cdot $ and the associated
internal energies as $E_{a}<E_{a^{\ast }}$, $\cdot \cdot \cdot $
and so on. \ \ Examples of $\psi _{int}$ and $E_{int}$  are
plotted in Fig.2 and listed in Table 1, respectively.

\vspace{1pt}\

\ \ \ \ \ Table 1, \ The lowest and second lowest internal energies (in $meV$%
)  of Type $a$ to $d$,\ $R=30nm.$

\ \ \

\begin{tabular}{|l|l|l|l|l|}
\hline
Type & $a$ & $b$ & $c$ & $d$ \\ \hline
$E_{int}$ & $2.626$ & $4.247$ & 2.630 & 4.272 \\ \hline
$E_{int}^{\ast }$ & $6.342$ & $8.912$ & 6.435 & 9.158 \\ \hline
\end{tabular}

\ \

\begin{figure} [htbp]
\centering
\includegraphics[totalheight=2.8in,trim=60 40 55 10]{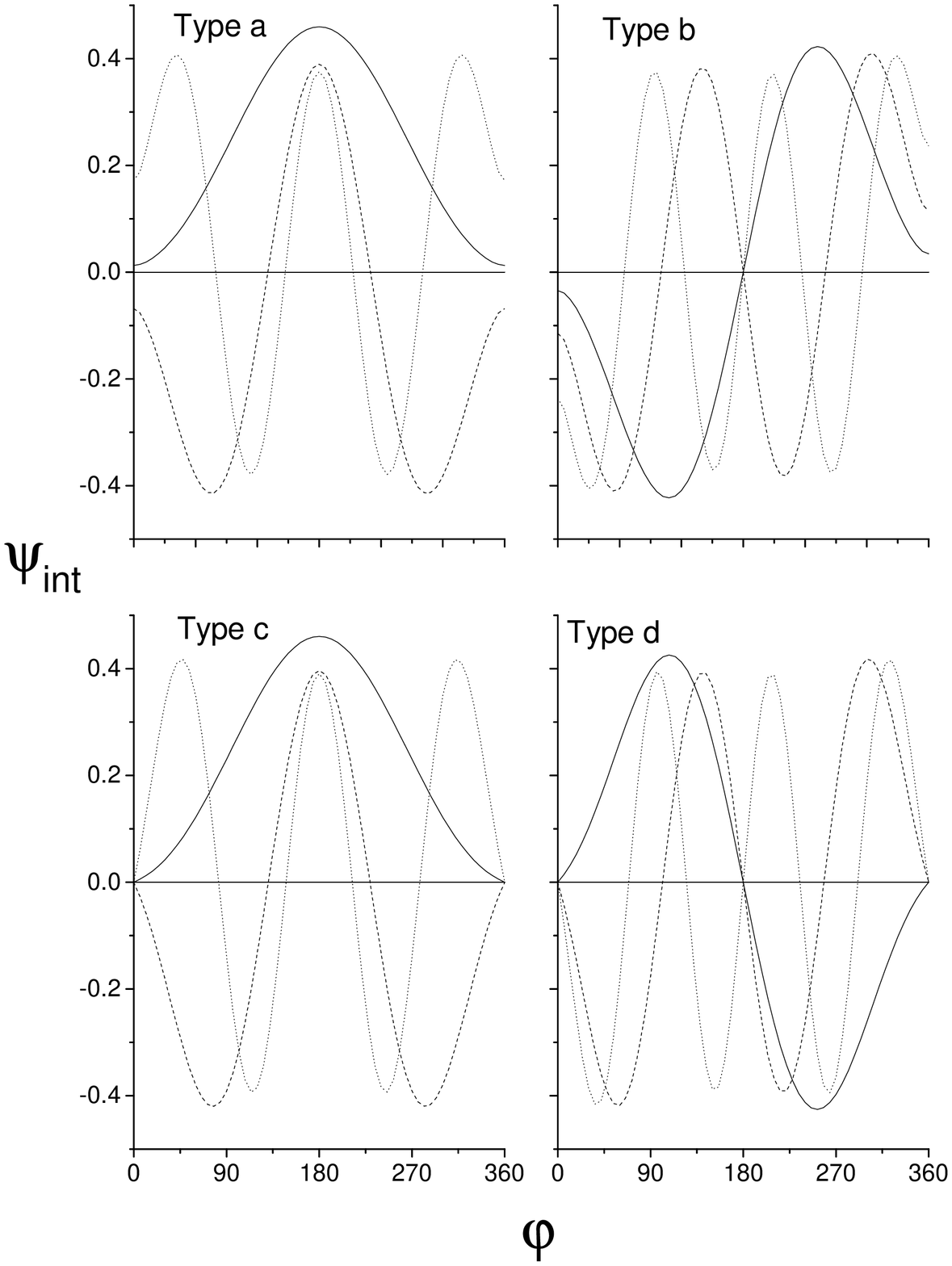}
\caption{Four types of $\psi _{int}$ against $%
\varphi $ , $R=40nm$. \ The lowest three of each type are shown,
the higher state has more nodes.} \label{Fig.2}
\end{figure}

 Due to the e-e repulsion, a dumbbell shape (DB), i.e., $\varphi
=180^{\circ }$, is advantageous in energy because the two
electrons are farther away from each other meanwhile. \ However, a
rotation\ of this geometry by $\pi $ is equivalent to an
interchange of particles, these operations will create the factors
$(-1)^{L}$\ and $(-1)^{S}$, respectively, from the wave function.
\ Therefore, the equivalence leads to a constraint, accordingly
the DB is allowed only for the states with $L+S$ even, \ i.e.,
only for Type $a$ and $c$. \ Otherwise, the states would have an
inherent node at the DB and therefore be higher in energy as shown
in Table 1, where $E_{a}<<E_{b}$, \ $E_{c}<<E_{d}$, and
$E_{a}\approx E_{c}$. \ \ In Fig.2 the patterns of Type $a$ are
one-to-one similar to Type $c$ , they all have a peak at the DB.
On the contrary, all those of Type ($b$) and($d$) have the
inherent node at the DB.  It is noticeable that Type $b$ and $c$
are not continuous at $\varphi =0$ and $2\pi $ due to their
periods are not equal to $2\pi $.  It was found that the internal
states of all the GSs are either $\psi _{a}$ or $\psi _{c}$\
without exceptions because the favorable DB is allowed in them. \
\ When the dynamical parameters vary in reasonable ranges, the
qualitative features of Fig.2 remain the same.

 According to (3), an appropriate $L_{o}$ would be chosen to minimize
the GS energy. \ When $\Phi $\ increases, $L_{o}$\ will undergo
even-odd transitions repeatedly and become more negative as shown
in Fig.1. Correspondingly, the total spin $S_{o}$ undergoes
singlet-triplet transitions, and $\psi _{a}$ and $\psi _{c}$\
appear in the GS alternatively. \ However, due to the Zeeman
effect, when $\Phi $ is larger than a critical value $\Phi
_{crit}$ , only $S_{o}=1$ states will be dominant, \ and
accordingly only $\psi _{c}$ will appear in the GS. \ The region
$\Phi <(>)\;\Phi _{crit}$ is called the FABO (ABO) region.

(iii) \textit{Persistent Current}:  Let $J_{1}$ be the current of
the particle $e_{1}$. The expression of $J_{1}$ is well known.$^5$
However, since it does not
depend on the azimuthal angle, \ it equals to its average over $\theta _{1}$%
. \ Thus the total current $J=J_{1}+J_{2}$ is

\ \ \ \

$J=\frac{1}{4\pi }g\int d\theta _{1}d\theta
_{2}\;\lbrack \Psi ^{\ast }(-i\frac{\partial }{\partial \theta _{1}}-i\frac{%
\partial }{\partial \theta _{2}}+2\Phi )\Psi +c.c.\rbrack \ \ \ \ \ (5)$

\ \ \ \

where $g=\hbar /(m^{\ast }R^{2})$. \ Using the arguments $\theta _{C}$ and $%
\varphi $ \ and making use of the separability, the integration over $\theta
_{C}$ and $\varphi $ can be performed. \ Thus we have

\ \ \ \

\qquad \qquad $J=g(L+2\Phi )/2\pi \qquad \qquad \qquad (6)$

\ \ \

This equation demonstrates explicitly the mechanism of the
oscillation of the persistent current, it is caused by the
step-by-step transition of $L$ during the increase of $\Phi $.
Examples of $J$ are shown in Fig.3, where each stronger
oscillation (associated with a $L$ odd and $S=1$ GS) is followed
by a weaker oscillation (associated with a $L$ even and $S=0$ GS).

\begin{figure} [htbp]
\centering
\includegraphics[totalheight=2.8in,trim=60 20 55 0]{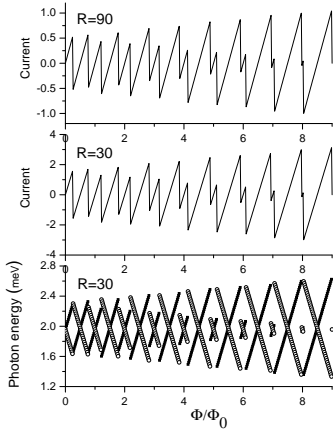}
\caption{The oscillation of the persistent current and the two
photon energies of the ground states against $\Phi /\Phi _{0}$ in
the FABO region. \ The unit of current is $10^{-5}C/R,$ where $C$
is the velocity of light. \ In the lowest panel, the black square
(white circle) denotes $\hbar \omega _{+}$ ($\hbar
\omega _{-}$), namely, the energy associated with $L_{o}$ to $L_{o}+1$ ($%
L_{o}-1$) transition.} \label{Fig.3}
\end{figure}

 (iv) \textit{Relations among the internal states}:  Define

  $O_{m}=e^{im(\theta _{1}-\theta _{C})}+$\ $e^{im(\theta
_{2}-\theta _{C})}=2\cos (m\varphi /2)$.  By analyzing the
numerical data, we found

\ \ \

 $\tilde{N}(O_{1}\psi _{a})=\psi _{b}+\xi _{a}$ and  $\tilde{N}(O_{1}\psi _{c})=\psi _{d}+\xi
_{c}\qquad (7)$

\ \ \ \

where $\tilde{N}$ is the operator of normalization, both $\xi
_{a}$ and $\xi _{c}$ are very small functions and depend on the
dynamical parameters very weakly. \ E.g., when $R$ varies from 30
to 90, the weights of $\xi _{a}$ and $\xi _{c}$ vary from $0.0004$
to $0.0002$. \ They are so small that in fact can be neglected.
Since $O_{1}$\ contains a node at the DB, it must cause a change
of type from $a$ to $b$, or from $c$ to $d$. Thus\ it is not
surprising that (7) holds. Since $O_{1}$\ is the operator of the
dipole transition (see below), eq.(7) provides an additional rule
of selection as discussed later.

(v) \textit{Dipole transition}:  The probability of dipole
transition reads $P_{(o),\pm }^{(f)}=\frac{ 2e^{2}}{3\hbar
}\;(\omega _{\pm }/c)^{3}R^{2}\;|A_{(o),\pm }^{(f)}|^{2}$, where
$\omega _{\pm }$\ is the frequency of the photon,

\ \ \ \

\qquad  $A_{(o)}^{(f)\pm }=\langle \Psi _{(f)\pm }|e^{\pm i\theta
_{1}}+e^{\pm i\theta _{2}}|\Psi _{(o)}\rangle$

\qquad \qquad \ \ \  $= \delta _{L_{(f)},\;L_{(o)}\;\pm 1}\langle
\psi _{int}^{(f)\pm }|O_{1}|\psi _{int}^{(o)}\rangle $ \qquad
\qquad \ (8)

\ \ \

where $(f)$ and $(o)$ denote the final and initial states,
respectively, the signs $\pm $ are associated with
$L_{(f)}=L_{(o)}\;\pm 1$.

 Let the initial state be the GS with $L_{o}$, then $\psi
_{int}^{(o)}$ must be $\psi _{a}$ or $\psi _{c}$ depending on
$L_{o}$\ is even or odd . Let $\alpha $ denotes the type of the
initial state. Due to (7) , $\langle \psi _{int}^{(f)\pm
}|O_{1}|\psi _{int}^{(o)}\rangle =\delta _{(f),\alpha }<O_{1}\psi
_{\alpha }|O_{1}\psi _{\alpha }>^{1/2}$, where $\delta _{(f)
,\alpha }$\ implies that the final state must be $\psi _b$ ($\psi
_ d$) if $\alpha =a$ ($c$), otherwise the amplitude is zero. Thus,
due to the additional rule of selection eq.(7), the dipole
strength of the GS is completely concentrated in two final states
having $L_{(f)}=L_{(o)}\;\pm 1$ and both having the same internal
state specified by eq.(7). \ Accordingly, only the photons with
the two energies

\ \ \ \

\qquad $\hbar \omega _{\pm }=E_{(f)\pm }-E_{(o)}$

\qquad \qquad $ =G\lbrack \frac{1}{2}(1\pm 2(L_{o}+2\Phi ))+\Delta
_{\alpha }/G\rbrack$ \qquad \ \ (9)

\ \ \ \

can be emitted (absorbed), where $\Delta _{\alpha }=E_{b}-E_{a}$ or $%
E_{d}-E_{c}$ depending on $\alpha =a$ or ($c$).\ \ The oscillation
of $\hbar \omega _{\pm }$ is plotted in the lowest panel of Fig.3.
It turns out that $\Delta _{\alpha }/G$ depends on $R$ very
weakly, thus $\hbar \omega _{\pm }$ is nearly proportional to
$R^{-2}$. Accordingly, a smaller ring will have a larger
probability of transition with a higher energy. \ \ \

(vi) \textit{FABO region}: The oscillation in this region is
complicated as shown in Fig.1 and 3. \ It is noted that the GS
energy (3), persistent
current (6), and the photon energies (9) all contain the factor $%
L_{o}+2\Phi ,$ thus their FABO are completely in phase and have
the same mechanism caused by the transition of $L_{o}$\ against
$\Phi $. \ \ In Fig.1 the abscissa $\Phi $ can be divided into
segments, \ in each the GS has a specific $L_{o}$\ and the GS
energy is given by a piece of a parabolic curve. \ The segment is
called an even (odd) segment if $L_{o}$\ is even (odd). \ At the
border of two neighboring segments  the two GS energies are equal.
\ From the equality and based on (3), the right and left
boundaries of the segment with $L_o$ can be obtained as

\ \ \ \

\qquad $\Phi _{right}(L_{o})=(1-(\mu /2G)^{2})^{-1}\lbrack \;1-\mu
(E_{c}-E_{a})/G^{2}-2L_{o}+(-1)^{L_{o}}(2(E_{c}-E_{a})+\mu
(L_{o}-1/2))/G\rbrack /4\qquad \qquad (10) $

\ \ \

$\Phi _{left}(L_{o})\;=(1-(\mu /2G)^{2})^{-1}\lbrack -1-\mu
(E_{c}-E_{a})/G^{2}-2L_{o}-(-1)^{L_{o}}(2(E_{c}-E_{a})+\mu
(L_{o}+1/2))/G\rbrack /4\qquad (11)$

\ \ \

where $L_{o}\leq 0$  and $\Phi _{right}(L_{o})=\Phi
_{left}(L_{o}-1)$, $\mu $ arises from the $H_{Zeeman}$. \ The
length of the segment reads

\ \ \ \

 $d_{L_{o}}=\Phi _{right}(L_{o})-\Phi _{left}(L_{o})=(1-(\mu
/2G)^{2})^{-1}\lbrack 1+(-1)^{L_{o}}(2(E_{c}-E_{a})+\mu
L_{o})/G\rbrack /2\qquad \qquad \qquad (12)$

which is related to the period of the FABO.  When $\Phi $\
increases, the magnitude of $L_o$ would increase.  Since $\mu L_o$
is negative, it is clear from eq.(12) that the length of even
(odd) segments would become shorter (longer) when $\Phi $\
increases.

The location of a segment with a given $L_{o}$ can be known from
the inequality $\Phi _{left}(L_{o})\leq \Phi \leq \Phi
_{right}(L_{o})$. \ Once the relation between $L_{o}$ and the
segments of $\Phi $ is clear, every details of the FABO can be
analytically and exactly explained via the eq.(3), (6), and (9).
In particular, the extrema in each segment can be known by giving
$\Phi =\Phi _{right}$ or $\Phi _{left}$. For an example, the
maximal current
 is $g(L_{o}+2\Phi _{right})/2\pi $.  Incidentally, the minimum of
 the GS energy in a segment is  $E_{\min }=E_{c}-\mu
^{2}/8G+\mu L_{o}/2$ (if $S_{o}=1$), or just equal to $E_{a}$ (if
$S_{0}=0$).

It is noted that $E_{c}-E_{a}$ (cf. Table 1) and $\mu /G$  (it is
$0.0554$ in our case) are both small. When $\Phi $\ is small the
magnitude of $|L_{o}|$ would be also small .  In this case eq.(12)
leads to $d_{L_{o}}\approx 1/2$, i.e., the period is a half of the
one of the normal ABO.  In fact, (12) provides an quantitative
description of the variation of the period of the FABO.

(vii) ABO \textit{region}:  When $\Phi $\ becomes sufficiently
large, $L_{o}$\ will become very
negative, \ the even segments will disappear due to their lengths $%
d_{L_{o}}\leq 0$ . We can define a critical odd integer $L_{crit}$\ so that $%
d_{L_{crit}-1}\leq 0$ \ while $d_{L_{crit}+1}>0$, thereby the critical flux
separating the FABO\ and ABO region can be defined as

\ \ \

\qquad $\Phi _{crit}=\Phi _{left}(L_{crit})\qquad \qquad (13)$

\ \ \

Once $\Phi >\Phi _{crit}$, $L_{o}$ remains odd and the system
keeps polarized. \ \ Let $I_{X}$ be the largest even integer
smaller than $-(G+2(E_{c}-E_{a}))/\mu $. \ It turns out from eq.(12) that $L_{crit}=I_{X}+1$%
. $\ $With our parameters, $L_{crit}=-19$ and accordingly $\Phi _{crit}=9.003
$\ ( refer to Fig.1). \ Both $L_{crit}$ and $\Phi _{crit}$ depend on $R$
very weakly, but sensitively on the effective mass $m^{\ast }$.\

In the ABO region ($\Phi >\Phi _{crit}$), eqs.(10) to (12) do not
hold. Instead we have $\Phi _{right}=-(L_{o}-1)/2,\;\Phi
_{left}=-(L_{o}+1)/2,$ and $d_{L_{o}}=1$. \ Thus the normal ABO
recovers. \
Evaluated from (6), the magnitude of current is from $-g/2\pi $ to $%
g/2\pi $ (for a comparison, it is from $-g/4\pi $ to $%
g/4\pi $ for 1-e rings). From (9) the photon energies $\hbar
\omega _{+}$ is from $\Delta _{c}-G/2$
 to $\Delta _{c}+3G/2$, \ at the same time $\hbar \omega _{-}$ is from $%
\Delta _{c}+3G/2$ \ to $\Delta _{c}-G/2$.

(viii) \textit{Relations between the photon energies and other
physical quantities}:  Due to (7),  the emitted (absorbed) dipole
photon has only two frequencies , therefore it is meaningful to
define $\Delta _{\hbar \omega }=\hbar (\omega _{+}-\omega _{-})$.
Directly from (9) and (6), we have

\ \ \

\qquad $\Delta _{\hbar \omega }=hJ_{o}\qquad \qquad (14)$

\ \ \

where $h$ is the Planck's constant and $J_{o}$\ is the persistent
current of the GS. To compare with 1-e rings, the latter has
$\Delta _{\hbar \omega }=2hJ_{o}$\. Eq.(14) demonstrates that the
oscillation of $\Delta _{\hbar \omega }$ and the
 oscillation of $J_{o}$ are matched with each other exactly,  they keep strictly
  proportional to each other during the variation of $\Phi $.

The maxima of $\Delta _{\hbar \omega }$ measured in the ABO and
FABO regions, respectively, read

\ \ \

\qquad $(\Delta _{\hbar \omega })_{\max }^{AB}=2G\qquad \qquad
(15)$

\ \ \

\qquad $(\Delta _{\hbar \omega })_{\max }^{FAB}=2G(L_{o}+2\Phi
_{right})\qquad \qquad (16)$

\ \ \

Obviously, (15) provides a way to determine $G$, $m^{\ast }$can be
thereby obtained. \ (16) can be rewritten as

\ \ \

$E_{c}-E_{a}=(G-\mu L_{o})/2-(2G-\mu )/(4G)(\Delta _{\hbar \omega
})_{\max }^{FAB}\ \ \ (17)$

This equation can be used to determine $E_{c}-E_{a}$.
Furthermore, we define

\ \ \

\qquad $\Gamma _{\hbar \omega }=\hbar (\omega _{+}+\omega
_{-})=G+2\Delta _{\alpha }\qquad \qquad (18)$

\ \ \

Once $G$\ has been known, (18) can be used to determine $E_{b}-E_{a}$ and $%
E_{d}-E_{c}$.  Since the spectrum can be generated from the
internal energies via (3), the evolutions of the spectrum and the
persistent current against $\Phi $\ can be understood simply by
measuring the photon energies.

\begin{figure} [htbp]
\centering
\includegraphics[totalheight=2.8in,trim=60 20 55 0]{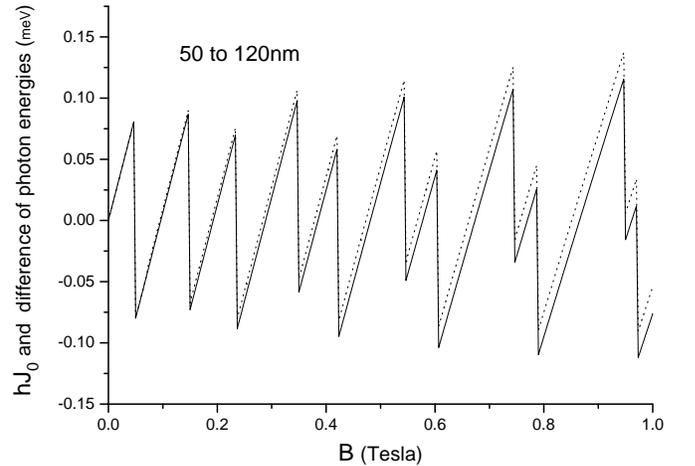}
\caption{Evolution of $hJ_{o}$(solid line) and $\Delta _{\hbar
\omega }$ (dotted line) against $B$ for a 2-e ring with $r_{a}=50$ and $%
r_{b}=120nm$} \label{Fig.4}
\end{figure}

(ix) \textit{Effect of the width}:  We now consider a
two-dimensional model in which the two electrons are strictly
confined in an annular region by a potential $U(r),$ which is zero if $%
r_{a}<r<r_{b}$ or is infinite otherwise. \ Under this model we
have performed numerical calculation to obtain $\Delta _{\hbar
\omega }$ and $hJ_{o}$, where $J_{o}$\ is now the total angular
current inside the ring (from $r_{a}$ to $r_{b}$). \ The result is
shown in Fig.4 where $r_{a}=50$ and $r_{b}=120$ are assumed, and
the two quantities are slightly different from each other. \
However, when the width becomes smaller, say $r_{b}-r_{a}<30$, the
two curves overlap. \ Thus (14) works not only for one-dimensional
but also for two-dimensional narrow rings. \ \ Let us define
$\stackrel{\_}{r}$\ $=\hbar /\sqrt{m^{\ast }(\Delta _{\hbar \omega
})_{\max }^{AB}}$.  For one-dimensional rings and from (15), we
have $\stackrel{\_}{r}=R$, where $R$ is the radius of the ring.
For two-dimensional rings, it was found from our numerical
calculation that  $\stackrel{\_}{r}$\ $\approx (r_{b}+r_{a})/2$ if
$r_{b}-r_{a}<30$.  E.g., when $r_{b}=100$ and $%
r_{a}=70$, $\stackrel{\_}{r}$\ =85.03. \ When $r_{b}=100$ and $r_{a}=90$, $%
\stackrel{\_}{r}$\ =95.00. \ \ Thus (15) works also well for
two-dimensional narrow rings if the $R$ in $G$\ is replaced by the
average radius.

It is noted that the band-structure and related optical properties
of 2-e rings have already been studied in detail by Wendler and
coauthors$^{18}$.  They classify the eigenstates according to
their radial motion, relative angular motion, and collective
rotation. In our paper the relative angular motion is further
classified into four types according to the inherent nodal
structures and periodicity of their wave functions, i.e.,
according to whether the DB shape is allowed and whether the wave
function is continuous at $\varphi =2\pi $.  The DB-accessibility
turns out to be important because it affects the eigenenergies
decisively.  In fact, the classification of states based on
inherent nodal structures was found to be crucial in atomic
physics,$^{19}$ this would be also true in two-dimensional
systems.  Furthermore, the rule of selection for the dipole
transition has been proposed in ref.[18].  In our paper, an
additional rule (namely,eq.(7)) is further proposed based on the
possible transition of internal structures.  This rule would
affect the dipole spectrum seriously because the emission
(absorption) is thereby concentrated into two frequencies.  The
difference of these two frequencies turns out to be proportional
to the persistent current.  Therefore the measurement of this
difference can be used to determine the magnitude of the current.

\qquad

In summary, we have studied the FABO both analytically and
numerically. The analytical formalism provides not only a base for
qualitative understanding, but also provides a number of formulae
for quantitative description. The domain of $\Phi $ is divided
into segments, each corresponds to a $L_o$.  This division
describes exactly how $L_o$ would transit against $|Phi $, which
causes directly the FABO. Thereby the variation of the period and
amplitude of the oscillation of the GS energy, persistent current,
and the frequencies of dipole transition in the FABO region can be
described exactly. A number of equalities to relate the physical
quantities and dynamical parameters have been found. In
particular, a new oscillation, namely, the oscillation of $\Delta
_{\hbar \omega }$ was found to match exactly the oscillation of
$J_o$. Since the photon energies can be more accurately measured,
other observables and parameters can be thereby determined via the
equalities. Since the separability of the Hamiltonian and the
existence of inherent nodes are common, the above description can be more or less generalized to $N-$%
electron rings, this deserves to be further studied. \

Acknowledgment, \ This work is supported by the NSFC of China under the
grants 10574163 and 90306016.

\qquad \qquad

.REFERENCES

1, A. Lorke, R.J. Luyken, A.O. Govorov, J.P. Kotthaus, J.M. Garcia, and P.M.
Petroff, \ Phys. Rev. Lett. 84, 2223 (2000).

2, U.F. Keyser, C. F\"{u}hner, S. Borck, R.J. Haug, M. Bichler, G.
Abstreiter, and W. Wegscheider, \ Phys. Rev. Lett. 90, 196601 (2003)

3, D. Mailly, C. Chapelier, and A. Benoit, \ Phys. Rev. Lett. 70, 2020 (1993)

4, A. Fuhrer, S. L\"{u}scher, T. Ihn, T. Heinzel, K. Ensslin, W.
Wegscheider, and M. Bichler, \ Nature (London) 413, 822 (2001)

5, S. Viefers, P. Koskinen, P.Singha Deo, M. Manninen, Physica E
21, 1(2004).

6, K. Niemel\"{a}, P. Pietil\"{a}inen, P. Hyv\"{o}nen, and T. Chakraborty, \
Europhys. Lett. 36, 533 (1996)

7, M. Korkusinski, P. Hawrylak, and M. Bayer, \ Phys. Stat. Sol. B 234, \
273 (2002)

8, Z. Barticevic, G. Fuster, and M. Pacheco, \ Phys. Rev. B 65, 193307 (2002)

9, M. Ferconi and G.Vignale, \ Phys. Rev. B 50, 14722 (1994).

10, \ Li. Serra, M. Barranco, A. Emperador, M. Pi, and E. Lipparini, \ Phys.
Rev. B 59, 15290 (1999)

11 \ A. Emperador, F. Pederiva, and E. Lipparini, \ Phys. Rev. B 68, 115312
(2003)

12, C.G. Bao, G.M. Huang, Y.M. Liu, \ \ Phys. Rev. B 72, 195310 (2005)\

13, A.E. Hansen, A. Kristensen, S. Pedersen, C.B. Sorensen, and
P.E. Lindelof, Physica E (Amsterdam) 12,770 (2002).

14, K. Moulopoulos and M. Constantinou, Phys. Rev. B. 70, 235327
(2004)

15,J. Planelles, J.I. Climente, and J.L. Movilla,
arXiv:cond-mat/0506691 (2005)

16, J.I. Climente and J. Planelles, Phys. Rev. B 72, 155322 (2005)

17, A.O. Govorov, S.E. Ulloa, K. Karrai, and R.J. Warburton, Phys.
Rev. B 66, 081309 (2002)

18, L. Wendler, V.M. Fomin, A.V. Chaplik, and A.O. Govorov, Phys.
Rev. B 54, 4794 (1996).

19, M.D. Poulsen and L.B. Madsen, Phys. Rev. A 72, 042501 (2005).

\vspace{1pt}

\end{document}